\documentclass[prd, aps, superscriptaddress, showpacs]{revtex4}

\begin{document}

\title{Quantum entanglement and entropy in particle creation}
\author{Shih-Yuin Lin}
\email{shihyuin@mail.ncku.edu.tw}
\email{sylin@cc.ncue.edu.tw}
\affiliation{Department of Physics, National Cheng Kung University and \\
Physics Division, National Center for Theoretical Sciences (South),
Tainan 70101, Taiwan}
\affiliation{Department of Physics, National Changhua University of Education,
Changhua 50007, Taiwan}
\author{Chung-Hsien Chou}
\email{chouch@mail.ncku.edu.tw}
\affiliation{Department of Physics, National Cheng Kung University and \\
Physics Division, National Center for Theoretical Sciences (South),
Tainan 70101, Taiwan}
\author{B. L. Hu}
\email{blhu@umd.edu}
\affiliation{Joint Quantum Institute and Maryland Center for Fundamental Physics,
University of Maryland, College Park, Maryland 20742-4111, USA}
\date{January 27, 2010}
%Final update: February 13, 2010
%First Notes: April 8, 2008

\begin{abstract}
We investigate the basic theoretical issues in the quantum entanglement of 
particle pairs created from the vacuum in a time-dependent background
field or spacetime. Similar to entropy generation from these processes 
which depends on the choice of physical variables and how certain information 
is coarse grained, entanglement dynamics hinges on the choice of measurable 
quantities and how the two parties are selected as well as the background 
dynamics of the field or spacetime. We discuss the conditions of separability 
of quantum states in particle creation processes and point out the differences 
in how the von Neumann entropy is used as a measure of entropy generation 
versus for entanglement dynamics. We show by an explicit construction that 
adoption of a different set of physical variables yields a different 
entanglement entropy. As an application of these theoretical considerations 
we show how the particle number and the quantum phase enter the entanglement 
dynamics in cosmological particle production.
\end{abstract}

 %Quantum Field Theory in Curved Spacetime

\maketitle

\section{Introduction}

Entanglement is said to be the uniquely distinguishing feature of ``quantumness"
\cite{Schrodinger}. Despite increasing interest in the recent decade and
advances in quantum information sciences, we are still far from fully
understanding the nature and dynamics of quantum entanglement  -- how it is
characterized and how it evolves in time -- in quantum open systems, i.e.,
those interacting with their environments. We believe at this stage of
development it is important to analyze these basic issues in great detail
in simple enough systems, preferably with exact solutions, so that we can
understand in depth its behavior to gain much needed valuable insight 
\cite{LH08, LCH08, LH09, LH2006, LHcqg}.

Usually entanglement is discussed in the framework of nonrelativistic
quantum mechanics, but for a quantity to have physical meaning one needs
to know how it transforms in different reference frames, e.g., for two
observers moving at relative constant speed how is the quantity one observer
reports as entanglement between two parties in its system related to that
reported by the other observer in its system? To answer this rather rudimentary
question one needs to work with Lorentz transformation of entanglement in the
framework of relativistic quantum mechanics (see, e.g., \cite{PerTer} for a
review).  This is the starting point of relativistic quantum information (RQI).
When a quantum field is involved, one needs to upgrade the treatment to that
of relativistic quantum field theory. This is the platform we have adopted for
our research program on RQI. When quantum informational issues arise pertaining
to black hole information loss or early universe quantum processes we need to
consider them in the extended framework of quantum field theory in curved
spacetime \cite{BirDav}.

The simplest process which distinguishes a quantum field theoretical process
from a quantum mechanical one is particle creation. The investigation of
quantum entanglement in the particle creation process -- how to define
entanglement, between what parties, and how it evolves in time -- is the first
order of business toward establishing a RQI theory for quantum field processes.
This is the goal of this paper, focusing on particle creation in strong and
dynamical background fields, such as in the Schwinger effect
\cite{Schwinger51}
and in background spacetimes, as in cosmological particle creation
\cite{Parker69,Zeldovich70}. In addition to its theoretical value these results
are expected to be  useful for quantum information experiments 
and for probes into the very early universe from next generation cosmological 
observations.

The statistical mechanical properties of particle creation such as entropy
generation have been a subject of both theoretical and cosmological interest for
quite some time. Since the mid-1980s there are inquiries on finding a viable
measure of entropy for particle creation processes in a free quantum field
\cite{HuPav86} and from interacting quantum fields \cite{HuKan87}. (The most
detailed discussion of entropy generation from free field particle creation 
along the theme of \cite{HuPav86} can be found in  \cite{KME98}. 
For a summary of recent developments see Chapter 9 of \cite{NEqFT08}.) 
The free field case is conceptually more intriguing: On the
one hand, from a pure quantum field theoretical perspective, one would say
categorically there could be no entropy generation because the particle pair
is originated from the vacuum which is a pure state. On the other hand, from
thermodynamical considerations one sees clearly that entropy is generated in the
amount proportional to the number of created particles.  This puzzle is what
started the inquiries mentioned above which led to the understanding that one
could quantify entropy generation by the number of particles created only because
one chooses to measure this process in the Fock (number) representation at the
sacrifice of the phase information, or effectively imposing a random phase
approximation. One also knows that it is only for the spontaneous production of
bosons that particle number increases monotonically, a necessary condition to
associate it with an entropy function obeying the second law. For fermions and
for stimulated creation processes particle number can decrease which invalidates
this definition. What this earlier investigation taught us is that the measured
entropy associated with the particle creation process depends on quite a few
factors: the choice of physical variables (such as using the number basis), the
coarse graining one introduces in its measurement (such as the quantum phase) or
recognizing that some  information is inaccessible to the observer in the process.

We mention prior studies on entropy generation in particle creation processes
because the experience we gained (concepts and methods) and the lessons we learned
(see above) prove to be useful for our investigations into the quantum entanglement
of particle creation because they share certain qualitative similarities.
Specifically, quantum entanglement depends crucially on the choice of physical
variables and the way the two parties whose entanglement is the object of
interest are defined. Quantum entanglement can be measured in many ways \cite{BZ06}. 
For the bipartite system the von Neumann (vN) entropy is quite commonly used. We 
will show that the way the vN entropy is used and what results it yields for these 
two processes, one pertaining to entanglement dynamics, the other for entropy
considerations in nonequilibrium statistical mechanics are quite different, both
in terms of whether and how coarse graining is introduced, and for what reasons.
A particle pair with some 3-momenta $({\bf k},-{\bf k})$ created from the vacuum 
is habitually regarded as perfectly entangled. We show by explicit construction 
that adoption of a different set of physical variables with the same 3-momenta 
$({\bf k}, -{\bf k})$ makes a difference in the entanglement dynamics. 

Before we go into detail, let us begin with a description of how information 
is chosen or coarse-grained in a closed versus an open quantum system. (For a 
treatment of entropy generation in squeezed open quantum systems with application 
to inflationary cosmology, see \cite{KMH} and related work cited therein.)
Any quantum state describing a closed (entirely isolated) system is always pure,
and the vN entropy is identically zero. In realistic settings a system is rarely
closed as it interacts, no matter how weakly, with its environments.
The back action of its environment with some coarse graining (in which
process noise is engendered) induces dissipation in the dynamics of the system,
now rendered open,  and mixed states are allowed in the open quantum system.
This is one way entropy is generated in the system.  Another way this could happen
is either by necessity, that some degree of imprecision always exists in
realistic measurements (or uncertainty in quantum state tomography), or by choice,
that only some physical variables of direct interests are measured (e.g., number
rather than phase in particle creation) and others ignored, averaged, or ``integrated
out" (e.g., the fast-oscillating elements of the density matrix). These procedures
either by necessity or by choice would render a pure state mixed in appearance
and entropy generation ensues.

The first type of entropy generation is well illustrated in an open quantum system
treatment of cosmological particle creation such as is shown in \cite{KMH}, and
more powerfully in the particle creation and backreaction problems, where the
gravitational sector is viewed as the (open) system and the quantum field as its
environment \cite{HuPhysica}, whose backreaction causes the dissipation of anisotropy 
or inhomogeneities \cite{HarHu79,CamVer94} of the early universe. One can
obtain the entropy generated in the particle creation process in terms of a vacuum
viscosity function, even define gravitational entropy \cite{Hu82} of
spacetime dynamics associated with these processes.  The second type is illustrated
clearly in free field particle creation processes \cite{HuPav86,KME98}, where the
choice of a Fock representation is what enables one to relate the amount of
entropy generated to the particle numbers created. Considering particle production
in a uniform electric field Kluger, Mottola, and Eisenberg (KME) \cite{KME98}
derived a quantum Vlasov equation describing the evolution of the (adiabatic) 
particle number, and an equation for the evolution of the quantum phase.  
They then argued that in the case they considered since the quantum phase as well 
as the off-diagonal elements of the density matrix (the coherence) 
oscillate rapidly in time, these fast variables are expected to be averaged out 
in observations and the density matrix would effectively look like a mixed state 
to such an observer.

Oftentimes these two sources of entropy generation are intermixed, depending on
how a problem is formulated and treated. This happens in the interacting field
particle creation calculations of \cite{HuKan87}. A more recent example is in
\cite{CP08, CP08b}, where Campo and Parentani 
studied the self-consistently truncated ``Gaussian and homogeneous density matrix" 
(GHDM) for an interacting field. They calculated the vN entropy of the two-mode 
sector with opposite wave vectors $({\bf k}, -{\bf k})$ of the GHDM and showed that 
the vN entropy is the only intrinsic property of the field state during inflation, 
while the entanglement between particles with ${\bf k}$ and their $-{\bf k}$ partners 
depends on the choice of canonical variables in the same mode pair $\phi_{\bf k}
\phi_{-\bf k}$. Their vN entropy of the two-mode sector plays a double role. 
On the one hand, it is a consequence of truncation or coarse graining in obtaining 
the GHDM, effectively by averaging out fast variables. On the other hand, it is a 
measure of the entanglement between that mode pair $\phi_{\bf k}\phi_{-\bf k}$ and its 
environment consisting of all other modes (rather than the entanglement of ${\bf k}$ 
and $-{\bf k}$ particles in the same two-mode sector), by noting that the GHDM is 
factorized and the two-mode sector with momenta $({\bf k}, -{\bf k})$ is by itself 
the reduced density matrix (at least approximately) with other degrees of freedom 
traced out.

To get rid of such intermixing, we are focusing on the simplest cases with free      %%%2/13
quantum fields in a dynamical background.                                            %%%
%The paper is organized as follows: 
In Secs. II and III we describe the particle
creation process of real and complex scalar fields, respectively, in the
Schr\"odinger representation. In Sec. II B and Sec. IV we investigate the 
behavior of particle numbers and quantum phase exploring the theoretical issues 
for the quantum entanglement of particle creation in a time-dependent background. 
Section II C contains the main results of entanglement dynamics using Wigner 
functions. After the theoretical issues are explored and analysis performed we 
study such processes in the early universe in Sec. V. We find that for the 
vacuum state of a free real scalar field in an expanding universe, once the 
physical variables are correctly chosen, it is possible to partition the degrees 
of freedom of a $({\bf k}, -{\bf k})$ mode pair into ${\bf k}$ and $-{\bf k}$ 
particles, and the degree of entanglement between them can be calculated 
accordingly. Based on these results we are able to look into how the particle 
number and the quantum phase enter the entanglement dynamics in cosmological 
particle production. We conclude in Sec. VI with a brief summary of the key
results reflecting the main themes stated here. 

%%%%%%%%%%%%%%%%%%%%%%%%%%%%%%%%%%%%%%%%%%%%%%%%%%%%%%%%%%%%%%%%%%%%%%%%%%
\section{Real scalar field with time-varying mass}
\label{PhiR}

Without loss of generality, let us work with a free scalar field in
Minkowski spacetime with a mass parameter $M(\eta)$ which is time dependent,
representing all time-dependent parameters entering into the system (including
that from cosmological spacetimes) 
\begin{equation}
  S = \int d^4 x \left[ -{1\over 2}\partial_\mu \Phi\partial^\mu\Phi -
    {1\over 2}M^2(\eta)\Phi^2 \right],
\end{equation}
where we denote time $x^0$ by  $\eta$ anticipating cosmological applications
later. In Fourier-transformed representation $\Phi(x)=(2\pi)^{-3}\int d^3k 
e^{i{\bf k\cdot x}}\phi^{}_{\bf k}$, the above action becomes
\begin{equation}
  S = \int d\eta {d^3k\over(2\pi)^3}\left[ {1\over 2}\partial_\eta \phi^{}_{\bf k}
    \partial_\eta\phi^{}_{-\bf k} -{1\over 2}\Omega^2_{\bf k}(\eta)\phi^{}_{\bf k}
    \phi^{}_{-\bf k}\right],
  \label{SphiRgen}
\end{equation}
where $\Omega^2_{\bf k}(\eta)=k^2+M^2(\eta)$ and $\phi^{}_{-\bf k}=\phi^*_{\bf k}$. 
The momentum conjugate to $\phi^{}_{\bf k}$ is $\Pi_{\bf k}=\delta S/\delta 
(\partial_\eta\phi_{\bf k}) = \partial_\eta\phi_{-\bf k}$, and the Hamiltonian 
can be derived straightforwardly by a Legendre transform of $S$.

In Sec. \ref{rsfFRW} we will see that the theory with a real scalar field in 
a Friedmann-Robertson-Walker (FRW) spacetime has the above form in conformal
time $\eta$.

%%%%%%%%%%%%%%%%%%%%%%%%%%%%%%%%%%%%%%%%%%%%%%%%%%%%%%%%%%%%%%%%%%%%%%%%%%%
\subsection{Quantization in Schr\"odinger representation}
\label{SchoPic}

%Consider the cases with $\kappa=0$.
Canonical quantization of this theory is achieved by imposing
the following equal-time commutation relations,
\begin{eqnarray}
  \left[ \phi^{}_{\bf k}(\eta ),\Pi_{{\bf p}}(\eta)\right] &=&
    i\hbar(2\pi)^3\delta^3({\bf k}-{\bf p}),\nonumber\\
  \left[ \phi^{}_{\bf k}(\eta ),\phi_{{\bf p}}(\eta)\right] &=&
    \left[ \Pi_{\bf k}(\eta ),\Pi_{{\bf p}}(\eta)\right]=0.\label{etcr}
\end{eqnarray}
In the Schr\"odinger representation \cite{GLH89}, $\phi^{}_{\bf k}$ are                 %%%2/13
viewed as c-number functions and the conjugate momentum operators are given by
\begin{equation}
  \hat{\Pi}_{\bf k} = (2\pi)^3{\hbar\over i}
    {\delta\over\delta\phi_{\bf k}},
\end{equation}
to satisfy the above commutation relations. The Hamiltonian operator then reads
\begin{equation}
  \hat{H} = \int {d^3k\over(2\pi)^3}\left[ -{\hbar^2 \over 2}(2\pi)^6
    {\delta\over\delta\phi_{\bf k}}{\delta\over\delta\phi_{-{\bf k}}} +
    {1\over 2}\Omega_{\bf k}^2(\eta)
    \phi^{}_{\bf k}\phi_{-{\bf k}} \right]. \label{ham}
\end{equation}
In this representation all physical information of the system is included
in the wave functional $\Psi [\phi^{}_{\bf k},\eta ]$ satisfying the
Schr\"odinger equation
\begin{equation}
  i\hbar\partial_\eta\Psi = \hat{H} \Psi . \label{Scheq}
\end{equation}

The ground state with minimum $\left<\right. \hat{H}\left.\right>$ is 
given by the Gaussian state
\begin{equation}
  \Psi_0 = s e^{-i \int^\eta d\bar{\eta} E_0(\bar{\eta})/\hbar}\exp
  -{1\over \hbar}\int {d^3k\over(2\pi)^3}\int {d^3p\over(2\pi)^3}
    \phi^{}_{\bf k} g^{\bf k ,p}(\eta) \phi^{}_{\bf p}, \label{ground}
\end{equation}
where $s$ is the normalization factor and $E_0$ is the vacuum energy
\footnote{In \cite{Be75}, Berger suggested an alternative choice of the
initial state $\Psi_B$ which satisfies
$(1/2)\Omega_{\bf k}(\eta_0)\Psi_B = \hat{H}(\eta_0)\Psi_B $
[which is stronger than $(\ref{Scheq})$] at the initial moment $\eta_0$.
She found that $\Psi_B = \sum_n a_{2n} \Psi_n$ with nonzero factors
$a_{2n}$ for $n=0,1,2,3,\ldots$. However, in her setup (Kasner universe),
$\Psi_B$ will become $\Psi_0$ as $\eta_0 \to -\infty$.}.
Substituting the above ansatz into $(\ref{Scheq})$, one finds that
\begin{equation}
  g^{\bf k,p}(\eta) = {1\over 2i} {{\chi^*_{\bf k}}'(\eta)\over
    \chi^*_{\bf k}(\eta)} (2\pi)^3\delta^3({\bf k} + {\bf p})
  \label{gfactor}
\end{equation}
so that $\Psi_0 = \prod_{\bf k} \Psi_{0\bf k}$, where
\begin{equation}
   \Psi_{0\bf k} = s_{\bf k}e^{-i \int^\eta d\bar{\eta}{\cal E}_0^{\bf k}
     (\bar{\eta})/\hbar}\exp {i\over 2\hbar(2\pi)^3\delta^3(0)}
     {{\chi^*_{\bf k}}'(\eta)\over \chi^*_{\bf k}(\eta)}
     \phi^{}_{\bf k}\phi^{}_{-\bf k}, \label{wavefnk}
\end{equation}
with the normalization factor $s_{\bf k}$ and the ``ground state energy"
for the ${\bf k}$ or $-{\bf k}$ modes,
\begin{equation}
  {\cal E}_0^{\bf k}\equiv {\hbar\over 2}\left[ |\chi'_{\bf k}|^2
      + \Omega_{\bf k}^2 |\chi^{}_{\bf k}|^2 \right],
  \label{E0k}
\end{equation}
such that $E_0(\eta)=\sum_{\bf k}{\cal E}_0^{\bf k}$. Here $\sum_k\equiv \int 
dk\,\delta(0)$, a prime denotes taking the derivative with respect to $\eta$,
and the mode function amplitudes $\chi^{}_{\bf k}$ satisfy the equation
\begin{equation}
  \chi''_{\bf k} +\Omega_{\rm k}^2(\eta) \chi^{}_{\bf k} = 0,
  \label{eomchi}
\end{equation}
so that $\chi^{}_{\bf k}=\chi^{}_{\bf -k}$.
This as we recognize is the classical field equation for the scalar field,
and $\chi^{}_{\bf k}$ is the solution for this field equation subject to the
normalization condition
\begin{equation}
  \chi^{}_{\bf k}\chi'^*_{\bf k} -\chi^*_{\bf k} \chi'_{\bf k}= i \label{wron}
\end{equation}
at every moment, which requires $\chi^{}_{\bf k}$ to be complex functions.
Note that $\chi^{}_{\bf k}$ does not have to be classically physical            %%% 2/13
solutions, because boundary conditions for $\chi^{}_{\bf k}$ will not be          %
fixed by classical considerations. In fact, different $\chi^{}_{\bf k}$         %%%
could correspond to different quantum states. Compared with the ground state 
for real scalar fields with constant mass in Minkowski space \cite{Hat92}, one 
sees that $\chi^{}_{\bf k}$ should be taken as $\sqrt{1/2\Omega_k} e^{-i
\Omega_k \eta}$ there, which is the positive frequency component of the 
${\bf k}$th mode of the free scalar field.

The wave functional in the form $(\ref{wavefnk})$ with $\chi_{\bf k}$              %%% 2/13
the solutions of the classical field equation is almost the same as an               %
infinite product of the wave functions for squeezed Gaussian states of 
parametric oscillators with frequency $\Omega_{\bf k}(\eta)$ \cite{KK99}, 
except that the variables in the exponents are mode pairs $\phi_{\bf k}
\phi_{\bf -k}$ for each $\bf k$ rather than the squared ``positions" of              %
the oscillators.                                                                   %%%

The nonvanishing two-point correlators in the vacuum state $(\ref{ground})$       %%% 2/13
are
\begin{equation}
  \left<\right.\phi^{}_{\bf k},\phi^{}_{\bf p}\left.\right> = \hbar
    (2\pi)^3\delta({\bf k}+{\bf p})\left| \chi^{}_{\bf k} \right|^2,
    \,\,\,\,\,
  \left<\right.\Pi_{\bf k},\Pi_{\bf p}\left.\right> = \hbar
    (2\pi)^3\delta({\bf k}+ {\bf p})\left| \chi'_{\bf k} \right|^2,
\end{equation}
and $\left<\right.\phi^{}_{\bf k},\Pi_{\bf p}\left.\right> = {1\over 2}
\partial_\eta \left<\right.\phi^{}_{\bf k},\phi^{}_{\bf p}\left.\right>$,
where $\left<\right. A,B\left.\right> \equiv {1\over 2}\int {\cal D}\phi
\Psi^*(AB+BA)\Psi$. Then one can write down the covariance matrix $V$ 
with elements $V_{ij}= \left<\right. {\cal R}_i, {\cal R}_j\left.\right>$ 
in ${\cal R}_i\equiv (\phi^{}_{\bf k},\Pi^{}_{\bf k},\phi^{}_{-\bf k},
\Pi^{}_{-\bf k})$ representation, whose partial transposition $V^{PT}
\equiv V|_{\Pi_{-\bf k} \to -\Pi_{-\bf k}}$ gives the quantity we used in 
an earlier paper \cite{LCH08, Si00}
\begin{eqnarray}
  \Sigma &\equiv& \det \left( V^{PT}_{ij}+{1\over 2}[{\cal R}_i, {\cal R}_j]
     \right) \nonumber\\
    &=& {1\over 4}\left[\hbar(2\pi)^3\delta^3(0)\right]^4 \left[ \left(
    \partial_\eta \left|\chi^{}_{\bf k}\right|^2 \right)^2+1\right].
\label{sigmaphiPi}
\end{eqnarray}
Since this is always positive, the field modes $\phi^{}_{\bf k}$ and
$\phi^{}_{-\bf k}$ are not only separable from other degrees of freedom
in the vacuum state (i.e., the von Neumann entropy of the density matrix
$\Psi^{}_{0\bf k}\Psi^*_{0\bf k}$ is zero, or in other words,
each $\Psi_{0\bf k}$ is always pure and factorizable from $\Psi_0$), but
also ``separable with each other" at all times. 
%This is consistent with the observation made in \cite{CP08}.

To see this more clearly, we write the complex $\phi^{}_{\bf k}$ in
terms of two real field variables as $\phi^{}_{\bf k}=\phi^R_{\bf k}+i
\phi^I_{\bf k}$ instead, with $\phi^R_{-\bf k}=\phi^R_{\bf k}$ and
$\phi^I_{-\bf k}= -\phi^I_{\bf k}$ (since $\phi^*_{\bf k}=\phi^{}_{-\bf k}$).
Then one can easily see that $(\ref{wavefnk})$ can be factorized into a
product of $\phi^R_{\bf k}$ state and $\phi^I_{\bf k}$ state. So if the
observables are ($\phi^R_{\bf k}, \phi^I_{\bf k}$) fields,
they will always be separable with each other and no entanglement measured 
in terms of these variables will be generated for a free scalar field
with time-varying mass, such as the field in an expanding universe.

Nevertheless, quantum entanglement depends on partition as well as the
choice of physical variables or measurables. With reference to quantum 
entanglement, say, in cosmology, foremost one needs to specify which 
physical observables are being measured there. Obviously $\phi^R_{\bf k}$ 
and $\phi^I_{\bf k}$ are not the correct variables to describe quantum
entanglement in cosmological particle creation: Recall that modern                %%% 2/13
cosmological experiments measure the temperature fluctuations $\delta T             %
/T$ of the cosmological microwave background radiation rather than the 
amplitudes of quantum fields. The former can be related to the 
energy density perturbation $\delta\rho/\rho$ with the energy density 
$\rho =\left<\right. {\cal H}\left.\right>\sim\sum_{\bf k}\left<\right.
N_{\bf k}\left.\right>/V$, where ${\cal H}$ and $N_{\bf k}$ are the
Hamiltonian density and the number operators for $\phi^{}_{\bf k}$,
respectively. %, which suggests that the field states in a Fock 
%representation associated with a particle number operator are those 
%physically measurable states. 
The number operator $N_{\bf k}$ here consists of creation and 
annihilation operators defining the in/out vacuum at the initial/final 
moment (or in the adiabatic vacuum). We therefore intend to look                    %
at quantum entanglement in terms of these operators.                              %%%
This also suggests that quantum entanglement generation, like cosmological 
particle creation, manifests only in those physical variables which 
facilitate a well-defined in/out or adiabatic vacuum.

%%%%%%%%%%%%%%%%%%%%%%%%%%%%%%%%%%%%%%%%%%%%%%%%%%%%%%%%%%%%%%%%%%%%%%%
\subsection{particle numbers}

Continuing our exposition using quantum field theory (QFT) in Minkowski space,
we define the annihilation and creation operators $b^{}_{\bf k}(\eta)$ and
$b^\dagger_{-\bf k}(\eta)$ by
\begin{eqnarray}
  b^{}_{\bf k}(\eta)&=& {-i\over\sqrt{\hbar(2\pi)^3\delta^3(0)}}\left(
  \chi'^*_{\bf k}(\eta)\phi^{}_{\bf k}-\chi^*_{\bf k}(\eta)\Pi_{-\bf k}\right),\\
  b^\dagger_{-\bf k}(\eta)&=& {i\over\sqrt{\hbar(2\pi)^3\delta^3(0)}}\left(
    \chi'_{\bf k}(\eta)\phi^{}_{\bf k}-\chi^{}_{\bf k}(\eta)\Pi^{}_{-\bf k}\right),
\end{eqnarray}
so that
\begin{eqnarray}
  \phi^{}_{\bf k}&=&\sqrt{\hbar(2\pi)^3 \delta^3(0)}
     \left(\chi^{}_{\bf k}(\eta)b^{}_{\bf k}(\eta)
    +\chi^*_{\bf k}(\eta) b^\dagger_{-\bf k}(\eta)\right),\label{phi2b}\\
  \Pi^{}_{-\bf k}&=&\sqrt{\hbar(2\pi)^3 \delta^3(0)}
     \left( \chi'_{\bf k}(\eta) b^{}_{\bf k}(\eta)+
     \chi'^*_{\bf k}(\eta)b^\dagger_{-\bf k}(\eta)\right),\label{Pi2b}
\end{eqnarray}
are independent of time. The above definition of operators $b^{}_{\bf k}(\eta)$
and $b^\dagger_{-\bf k}(\eta)$ has the following properties: First, they
become the conventional ones in QFT in Minkowski space where $\chi^{}_{\bf k}=
\sqrt{1/2\Omega_k} e^{-i\Omega_k \eta}$. Second, the equal-time commutation 
relations Eq.$(\ref{etcr})$ are equivalent to
\begin{equation}
  [b^{}_{\bf k}, b^\dagger_{\bf p}]=\delta^3({\bf k}-{\bf p})/\delta^3(0).
\end{equation}
Third, the ground state $\Psi_0$ is the state with the property $b_{\bf k}
\Psi_0=0$ for all ${\bf k}$. Fourth, the ``excited states" are analogous to 
those in simple harmonic oscillators, which are generated by applying these
creation and annihilation operators to the ground states, e.g.,
\begin{eqnarray}
  \Psi_1({\bf k})&=& b^\dagger_{-\bf k}\Psi_0
  ={1\over\sqrt{\hbar(2\pi)^3\delta^3(0)}}{\phi^{}_{\bf k}\over\chi^*_{\bf k}}\Psi_0,\\
  \Psi_2({\bf k},-{\bf k})&=& b^\dagger_{\bf k}b^\dagger_{-\bf k}\Psi_0 =
    \left({1\over\hbar(2\pi)^3\delta^3(0)}
    {\phi^{}_{\bf k}\phi^{}_{-\bf k}\over\chi^*_{\bf k}\chi^*_{-\bf k}} -
    {\chi^{}_{\bf k}\over \chi^*_{\bf k}}\right)\Psi_0  , \, etc,
\end{eqnarray}
are also solutions of the Schr\"odinger equation.
Moreover, all excited states generated in this way, together with the ground
state, form a complete set of quantum states for the scalar field.
%To be proven.

Note that a proper normalization for $b^{}_{\bf k}$ and $b^\dagger_{-\bf k}$
has been chosen to make the above excited states satisfy the same
normalization conditions for $\Psi_0$. Written in terms of $b^{}_{\bf k}$ and
$b^\dagger_{-\bf k}$, the Hamitonian operator reads
\begin{equation}
  \hat{H}=\sum_{\bf k}
    %\int d^3k 
    \left\{ {\cal E}_0^{\bf k}\left(
    b^{}_{\bf k}b^\dagger_{\bf k}+ b^\dagger_{\bf k}b^{}_{\bf k}\right) +
    {\hbar\delta^3(0)\over 2}\left[\left( \chi'_{\bf k}{}^2+
    \Omega_{\bf k}^2 {\chi^{}_{\bf k}}^2\right)
    b^{}_{\bf k}b^{}_{\bf -k}+{\rm H.c.}\right]\right\},
\end{equation}
where ``H.c." stands for ``Hermitian conjugate" and ${\cal E}_0^{\bf k}$
is the ${\bf k}$th component of the ground state energy given in $(\ref{E0k}
)$. So $\int {\cal D}\phi\,\Psi^*_0(\eta) \hat{H}\Psi^{}_0(\eta)= E_0(\eta)$
is indeed the vacuum energy.

The ``number operator" of mode $\bf k$ at the moment $\eta$ is defined as
\begin{equation}
  \hat{N}_{\bf k}(\eta)\equiv b_{\bf k}^\dagger (\eta )b^{}_{\bf k}(\eta).
  \label{PNOp}
\end{equation}
Suppose at the initial moment $\eta_0$ the field is in the vacuum state when
the particle-number counter is constructed according to the operator
$\hat{N}_{\bf k}(\eta_0)$. Then, using the same particle-number counter,
one finds that the number of the particle created at time $\eta$ by
the background spacetime is
\begin{equation}
  \left< 0_\eta \right| \hat{N}_{\bf k}(\eta_0)\left| 0_\eta\right> \equiv
  \int {\cal D}\phi\,\Psi^*_0(\eta) \hat{N}_{\bf k}(\eta_0)\Psi^{}_0(\eta) = 
  \left| \chi'_{\bf k}(\eta_0 )\chi^{}_{\bf k}(\eta )-
     \chi^{}_{\bf k}(\eta_0 )\chi'_{\bf k}(\eta )\right|^2 .
\label{PartNo}
\end{equation}
Since both $\chi^{}_{\bf k}(\eta_0 )$ and $\chi^{}_{\bf k}(\eta )$ are solutions
of $(\ref{eomchi})$ and each form a complete set at different times, one
may write
\begin{equation}
  \chi^{}_{\bf k}(\eta_0)=
  \alpha^{}_{\bf k}(\eta_0,\eta)\chi^{}_{\bf k}(\eta)
  +\beta^{}_{\bf k}(\eta_0,\eta)\chi^*_{\bf k}(\eta), \label{chi0AB}
\end{equation}
or equivalently, for $|\alpha^{}_{\bf k}|^2-|\beta^{}_{\bf k}|^2=1$,
one has
\begin{equation}
  \chi^{}_{\bf k}(\eta)=
  \alpha^*_{\bf k}\chi^{}_{\bf k}(\eta_0)
  -\beta^{}_{\bf k}\chi^*_{\bf k}(\eta_0). \label{chiAsB}
\end{equation}
Then the Bogoliubov coefficients read
\begin{eqnarray}
  \alpha^{}_{\bf k} &=&
      i\left[ \dot{\chi}^{}_{\bf k}(\eta_0)\chi^*_{\bf k}(\eta)
     -\chi^{}_{\bf k}(\eta_0) \chi'^*_{\bf k}(\eta)\right] +
      i\zeta(\eta)\chi^*_{\bf k}(\eta)\chi^{}_{\bf k}(\eta_0) ,
    \label{bogoA}\\
  \beta^{}_{\bf k} &=&
      i\left[\chi^{}_{\bf k}(\eta_0)\chi'_{\bf k}(\eta) -
          \dot{\chi}^{}_{\bf k}(\eta_0)\chi^{}_{\bf k}(\eta)\right] -
      i\zeta(\eta)\chi^{}_{\bf k}(\eta)\chi^{}_{\bf k}(\eta_0) ,
    \label{bogoB}
\end{eqnarray}
with overdots denoting $\partial/\partial \eta |_{\eta=\eta_0}$ and
$\zeta \in {\bf R}$, whose values will be further fixed by specifying
$\chi'_{\bf k}$ \cite{KME98} or other physical conditions.
From $(\ref{PartNo})$ it can be seen that we have made the physical choice
$\zeta=0$, which implies $\left< 0_\eta \right|\hat{N}_{\bf k}(\eta_0)
\left|0_\eta\right> = |\beta_{\bf k}|^2$.

Note we did not get into the details about how to define an adiabatic 
particle-number state (which can be found in e.g., \cite{KME98, BirDav}) but 
simply assume that at $\eta_0$ the concept of particle is well defined. 
Here the Bogolubov coefficients with $\zeta=0$ are represented explicitly 
by the Klein-Gordon inner product.

%%%%%%%%%%%%%%%%%%%%%%%%%%%%%%%%%%%%%%%%%%%%%%%%%%%%%%%%%%%%%%%%%%%%%%%%%%
\subsection{Wigner function and entanglement entropy}

Since $\Psi_0$ is a pure state, if one divides the degrees of freedom
in this model into two parties, then the entanglement between them
can be well measured by the von Neumann entropy of the reduced density
matrix of one of the two parties. %, or the entanglement entropy.
Nevertheless, it is not obvious whether the particles with
${\bf k}$ and $-{\bf k}$ are separable as indicated by the positive
$\Sigma$ at the end of Sec.\ref{SchoPic}.
%The fact that the particle number $N_{\bf k}$ is defined by operators
%in Fock space representation suggests that we could see cosmological
%particle creation more clearly by looking into the Wigner function in
%Fock space representation.

A clearer separability can be seen in the Wigner function of $\Psi_0$:
\begin{eqnarray}
  \rho(\eta) &=& \int \prod_{{\bf k} \in ({\bf R}^3 -\{ 0\})/{\bf Z}_2}
  \left[ d\Delta^R_{\bf k}
  d\Delta^I_{\bf k} e^{i\left(p^R_{\bf k} \Delta^R_{\bf k}+ 
    p^I_{\bf k}\Delta^I_{\bf k}\right)/\hbar} \Psi_{0\bf k}\left(
          \phi^R_{\bf k} -{\Delta^R_{\bf k}\over 2},
      \phi^I_{\bf k} -{\Delta^I_{\bf k}\over 2}\right)
    \Psi_{0\bf k}^*\left(
      \phi^R_{\bf k} +{\Delta^R_{\bf k}\over 2},
      \phi^I_{\bf k} +{\Delta^I_{\bf k}\over 2}\right)\right]\nonumber\\
  &=& |\tilde{s}|^2 \exp -{2\over \hbar}\int {d^3 k\over (2\pi)^3}\left\{
  \left| \chi'_{\bf k}(\eta)\right|^2\phi^{}_{\bf k}\phi^{}_{-\bf k} +
  \left| \chi^{}_{\bf k}(\eta)\right|^2 \pi^{}_{\bf k}\pi^{}_{-\bf k}
  -\partial_\eta\left( \left| \chi^{}_{\bf k}(\eta)\right|^2\right)
  \pi^{}_{\bf k}\phi^{}_{-\bf k} \right\},
\end{eqnarray}
where we have taken
$\phi^{}_{\bf k}=\phi^R_{\bf k}+i\phi^I_{\bf k}$ (with $\phi^R_{-\bf k}=
\phi^R_{\bf k}$, $\phi^I_{-\bf k}=-\phi^I_{\bf k}$) and $\pi_{\pm \bf k}
= (p^R_{\bf k}\pm i p^I_{\bf k})/2$, which are c-numbers here rather
than operators.
%, to make the corresponding commutation relations consistent.
Performing a canonical transformation similar to $(\ref{phi2b})$ and
$(\ref{Pi2b})$ as
\begin{eqnarray}
  \phi^{}_{\bf k}&=&\sqrt{\hbar(2\pi)^3\delta^3(0)}\left(
      \chi^{}_{\bf k}(\eta)\tilde{B}^{}_{\bf k}(\eta) +
      \chi^*_{\bf k}(\eta) \tilde{B}^*_{-\bf k}(\eta)\right),\label{phiB}\\
  \pi^{}_{\bf k}&=&\sqrt{\hbar(2\pi)^3\delta^3(0)}\left(
      \chi'_{\bf k}(\eta)\tilde{B}^{}_{\bf k}(\eta)+
      \chi'^*_{\bf k}(\eta)\tilde{B}^*_{-\bf k}(\eta)\right),\label{PiB}
\end{eqnarray}
which gives $d\phi^{}_{\bf k}d\pi^{}_{\bf k}=d\tilde{B}^{}_{\bf k}d
\tilde{B}^*_{\bf k}\hbar(2\pi)^3\delta^3(0)$, one ends up with the Wigner 
function in a Fock representation,
\begin{equation}
  \rho(\eta) = |\tilde{s}|^2 \exp -2\sum_{\bf k}\tilde{B}^{}_{\bf k}(\eta)
    \tilde{B}^*_{\bf k}(\eta). \label{WignerBt}
\end{equation}
Here one can easily see that the quantum states of $(\tilde{B}^{}_{\bf k},
\tilde{B}_{\bf k}^*)$ and $(\tilde{B}^{}_{-\bf k},\tilde{B}_{-\bf k}^*)$
for each specific ${\bf k}$ are separable.

However, in terms of an alternative set of measurables,
${\bf k}$ and ${-\bf k}$ particles could be entangled
as we will see in the following. From $(\ref{PartNo})$ it seems that the
physical particle counter should be counting $\hat{N}_{\bf k}(\eta_0)$ 
defined by operators $b_{\bf k}$ at the initial moment $\eta_0$. Let 
$B^{}_{\bf k}\equiv \tilde{B}^{}_{\bf k}(\eta_0)$ in $(\ref{phiB})$ and 
$(\ref{PiB})$ at $\eta=\eta_0$, so that
\begin{equation}
  \tilde{B}_{\bf k}(\eta) = \alpha^{}_{\bf k} B^{}_{\bf k} +
    \beta^*_{\bf k} B^*_{-\bf k}, \label{BnBt}
\end{equation}
from $(\ref{chiAsB})$. Then in terms of $B^{}_{\bf k}$ with $n_k \equiv
|\beta_{\bf k}|^2$, $c_{\bf k}\equiv\alpha_{\bf k} \beta_{\bf k}$, the
Wigner function $(\ref{WignerBt})$ becomes
\begin{equation}
  \rho(\eta) = |\tilde{s}|^2 \exp -2\sum_{\bf k}\left[
    \left(2n_{\bf k}+1\right)B^{}_{\bf k} B_{\bf k}^* +
    c^{}_{\bf k} B^{}_{\bf k} B^{}_{-\bf k} +
    c^*_{\bf k}B_{\bf k}^* B_{-\bf k}^*\right],
  \label{WignerB}
\end{equation}
which gives the ``symmetric" particle number
\begin{equation}
  \left<\right. B_{\bf k}^* B^{}_{\bf k} \left.\right> =
  n_{\bf k}+{1\over 2} = \left< 0_\eta\right|b_{\bf k}^\dagger(\eta_0),
  b^{}_{\bf k}(\eta_0)\left| 0_\eta \right>. 
\end{equation}
This justifies $B^{}_{\bf k}$ as the correct variables corresponding to
physical measurements. Suppose we divide the particles into two groups
with $k^3>0$ and $k^3<0$ respectively (that the particles with $k^3=0$ are 
properly divided into these two groups is understood). One may write
\begin{equation}
  \rho(\eta) = |\tilde{s}|^2 \exp -{\sum_{\bf k}}^+ \left[
  \left(4n_{\bf k}+2\right)
  \left(B^{}_{\bf k} B_{\bf k}^* + B^{}_{-\bf k} B_{-\bf k}^*\right)
  + 4c^{}_{\bf k}B^{}_{\bf k}B^{}_{-\bf k}
  + 4c^*_{\bf k}B_{\bf k}^*B_{-\bf k}^*\right].
\end{equation}
where ${\sum_{\bf k}}^+$ and later ${\prod_{\bf k}}^+$ denote
summing and multiplying over ${\bf k}$ with $k^1$, $k^2 \in {\bf R}$
and $k^3>0$, respectively. Integrating out $B_{-\bf k}$ and $B_{-\bf k}^*$ in
$\rho(\eta)$, one obtains the reduced Wigner function
\begin{eqnarray}
  \rho^R(\eta) &\equiv& \int {\prod_{\bf k}}^+
    dB^{}_{-\bf k} dB_{-\bf k}^* \rho(\eta)\nonumber\\
  &=& |\bar{s}|^2 \exp -{\sum_{\bf k}}^+ {2 B^{}_{\bf k} B_{\bf k}^* \over
    2n_{\bf k}+1}\left[ \left(2n_{\bf k}+1\right)^2-
    4\left|c_{\bf k}\right|^2\right]\nonumber\\
  &=& |\bar{s}|^2 \exp -{\sum_{\bf k}}^+
    {2 B^{}_{\bf k} B_{\bf k}^* \over 2n_{\bf k}+1},
  \label{RedWignerB}
\end{eqnarray}
which is a mixed state once $|\beta_{\bf k}|\not= 0$ because the purity
\begin{equation}
  {\cal P} = 2\pi \int dB^{}_{\bf k} dB^*_{\bf k} \left(\rho^R\right)^2
  = {\prod_{\bf k}}^+ {1\over 2n_{\bf k} +1} \label{pur}
\end{equation}
is less than $1$ if any $n_{\bf k}=|\beta_{\bf k}|^2 > 0$. This means that
the particles measured by ``physical particle counter" $\hat{N}_{\bf k}
(\eta_0)$ with ${\bf k}$ and $-{\bf k}$ are entangled. Comparing the above
expression for the purity with Eqs. (30) and (31) of Ref.\cite{LH2006},
one sees that the von Neumann entropy is
\begin{eqnarray}
  {\cal S} &=& {\sum_{\bf k}}^+ \left[(n_{\bf k}+1)\ln (n_{\bf k}+1) -
    n_{\bf k}\ln n_{\bf k}\right] \nonumber\\
    &=& {1\over 2}{\sum_{\bf k}} \left[(n_{\bf k}+1)\ln (n_{\bf k}+1) -
    n_{\bf k}\ln n_{\bf k}\right].  \label{vNEn}
\end{eqnarray}
When $n_{\bf k}\gg 1$ for all ${\bf k}$, ${\cal S}\approx 
{\sum_{\bf k}}^+ \ln n_{\bf k}$.

vN entropy as an entanglement entropy measures the quantum nonlocal             %%% 2/13
correlations at some moment between the system and some specified party 
which could be the rest of the world. The latter is traced out at the moment 
the entanglement entropy is evaluated. Note that ``nonlocal" here does not 
imply ``nonlocal in space": the particles in this paper are actually 
something similar to plane waves, which are local in momentum                   %%% 2/13
space but nonlocal in position space. They are not spacelike                      %
separated. Rather, they can occupy the same position space at the same time.    %%%

%%%%%%%%%%%%%%%%%%%%%%%%%%%%%%%%%%%%%%%%%%%%%%%%%%%%%%%%%%%%%%%%%%%%%%%%%%
\section{Complex scalar field with time-varying mass}
\label{PhiC}

One can easily generate the above formulations to a complex field
with time-varying mass:
\begin{equation}
  S = \int d^4 x \left[ -{1\over 2}\partial_\mu \Phi\partial^\mu\Phi^* -
    {1\over 2}M^2(\eta)|\Phi|^2 \right],
\end{equation}
or in a Fourier-transformed representation,
\begin{equation}
  S = \int d\eta {d^3k\over(2\pi)^3}\left[ {1\over 2}\partial_\eta
    \phi^{}_{\bf k}\partial_\eta\phi^*_{\bf k} -
    {1\over 2}\Omega^2_{\bf k}(\eta)|\phi^{}_{\bf k}|^2\right],
  \label{SphiCgen}
\end{equation}
The only difference is that now $\phi^*_{\bf k}$ is independent of
$\phi^{}_{-\bf k}$, while $\Pi^{}_{\bf k} = \partial_\eta \phi^*_{\bf k}$.       %%% 2/13
The above action describes scalar QED of a quantum charged scalar
field in a classical uniform electric field \cite{KME98}.

The ground state wave function is still in the form of $(\ref{wavefnk})$,
except that the $\phi^{}_{-\bf k}$ should be replaced by $\phi^*_{\bf k}$.
But now the particles and antiparticles can be distinguished by their
different charges in addition to their momenta. So instead of
$(\ref{phi2b})$ and $(\ref{Pi2b})$, one writes
\begin{eqnarray}
  \phi^{}_{\bf k}&=&\sqrt{\hbar(2\pi)^3 \delta^3(0)}
     \left(\chi^{}_{\bf k}(\eta)a^{}_{\bf k}(\eta)
    +\chi^*_{\bf k}(\eta) b^\dagger_{-\bf k}(\eta)\right),\label{phi2ab}\\
  \Pi^*_{\bf k}&=&\sqrt{\hbar(2\pi)^3 \delta^3(0)}
     \left( \chi'_{\bf k}(\eta) a^{}_{\bf k}(\eta)+
     \chi'^*_{\bf k}(\eta)b^\dagger_{-\bf k}(\eta)\right),\label{Pi2ab}
\end{eqnarray}
with
\begin{equation}
  [a^{}_{\bf k}, a^\dagger_{\bf p}]=
  [b^{}_{\bf k}, b^\dagger_{\bf p}]=\delta^3({\bf k}-{\bf p})/\delta^3(0).
\end{equation}
This will make the Wigner function of the vacuum state a product
of two copies of $(\ref{WignerBt})$:
\begin{equation}
  \rho(\eta) = |\tilde{s}|^2 \exp -2\sum_{\bf k}\left(
    \tilde{A}^{}_{\bf k}(\eta)\tilde{A}_{\bf k}^*(\eta)+
    \tilde{B}^{}_{\bf k}(\eta)\tilde{B}_{\bf k}^*(\eta)\right),
  \label{rhoAtBt}
\end{equation}
where $(\tilde{A}^{}_{\bf k}, \tilde{B}^{}_{\bf k})$ corresponding to
$(a^{}_{\bf k}, b^{}_{\bf k})$ are the counterpart of $\tilde{B}^{}_{\bf k}$
of the real scalar field. Thus as before one can express them in terms
of the ones defined at the initial moment as
%\end{equation}
\begin{equation}
  \tilde{A}^{}_{\bf k}(\eta) = \alpha^{}_{\bf k} A^{}_{\bf k} +
    \beta^*_{\bf k} B^*_{-\bf k}, \,\,\,
  \tilde{B}^{}_{-\bf k}(\eta) = \beta^{}_{\bf k} A^{}_{\bf k} +
    \alpha^*_{\bf k} B^*_{-\bf k},
\end{equation}
according to $(\ref{chi0AB})$. These imply
\begin{equation}
  \rho(\eta) = |\tilde{s}|^2 \exp -2{\sum_{\bf k}} \left[
    \left(2n_{\bf k}+1\right) \left(A^{}_{\bf k} A_{\bf k}^*+
    B^{}_{\bf k} B_{\bf k}^*\right) + c^{}_{\bf k}
    (A^{}_{\bf k}B^{}_{-\bf k}+A^{}_{-\bf k}B^{}_{\bf k}) + c^*_{\bf k}
    (A_{\bf k}^* B_{-\bf k}^*+A_{-\bf k}^* B_{\bf k}^*)\right].
\label{rhoAB}
\end{equation}
Although a mixing between the particles with ${\bf k}$ and the
antiparticles with $-{\bf k}$ is generated, the vacuum state
$\rho(\eta)$ remains a pure state. Only after one integrates out the
antiparticles (particles) associated with $B^{}_{\bf k}$ ($A^{}_{\bf k}$) 
will the reduced Wigner function for particles (antiparticles),
$\rho^A$ ($\rho^B$), become
\begin{equation}
  \rho^C(\eta) = |\bar{s}|^2 \exp -{\sum_{\bf k}}
    {2 C^{}_{\bf k} C_{\bf k}^* \over 2n_{\bf k}+1},
\label{redrhoC}
\end{equation}
with $C = A,B$. Now $\rho^C$ is a mixed state. The purity and the von
Neumann entropy are those for the real scalar field $(\ref{pur})$ and
$(\ref{vNEn})$ with $\prod^+_{\bf k}$ and $\sum^+_{\bf k}$ replaced by
the usual $\prod_{\bf k}$ and $\sum_{\bf k}$. So the values of the von
Neumann entropy of a complex scalar field between particles and
antiparticles are twice the value for a real scalar
field, indicated by $(\ref{vNEn})$.

%%%%%%%%%%%%%%%%%%%%%%%%%%%%%%%%%%%%%%%%%%%%%%%%%%%%%%%%%%%%%%%%%%%%%%%%
\section{Phase information}
\label{phaseinf}

Observe that the description from $(\ref{WignerBt})$ to $(\ref{WignerB})$
is that of squeezing in a two-mode squeezed state, well known from a
squeezed-state description of particle creation (see, e.g., \cite{HKM,KMH}).
Writing $\tilde{B}_{\bf k}$ in terms of quadrature amplitudes, namely,
$\tilde{B}_{\bf k}= (\tilde{Q}_{\bf k}+ i \tilde{P}_{\bf k})/\sqrt{2}$
with $\tilde{Q}_{\bf k}$ and $\tilde{P}_{\bf k}$ real, then
$(\ref{WignerBt})$ looks like a direct product of the Wigner functions
for the ground states of $\tilde{Q}_{\bf k}$ for all ${\bf k}$. Now,
since $|\alpha_{\bf k}|^2 -|\beta_{\bf k}|^2 =1$, one is allowed to
parametrize $\alpha_{\bf k}$ and $\beta_{\bf k}$ as
\begin{equation}
  \alpha_{\bf k} \equiv e^{i\sigma_{\bf k}}\cosh r_{\bf k}, \,\,\,\,\,
  \beta_{\bf k} \equiv -e^{i\theta_{\bf k}}\sinh r_{\bf k}.
  \label{squeezpara}
\end{equation}
Let $B^{}_{\bf k}\equiv(Q_{\bf k}+ i P_{\bf k})/\sqrt{2}$. Then one has
\begin{eqnarray}
  Q_{\bf k} &=& \cosh r_{\bf k} \left(\tilde{Q}_{\bf k}\cos\sigma_{\bf k}
    +\tilde{P}_{\bf k}\sin\sigma_{\bf k}\right) +
    \sinh r_{\bf k} \left(\tilde{Q}_{-\bf k}\cos\theta_{\bf k}
    -\tilde{P}_{-\bf k}\sin\theta_{\bf k}\right),\\
  P_{\bf k} &=& \cosh r_{\bf k} \left(-\tilde{Q}_{\bf k}\sin\sigma_{\bf k}
    +\tilde{P}_{\bf k}\cos\sigma_{\bf k}\right) -
    \sinh r_{\bf k} \left(\tilde{Q}_{-\bf k}\sin\theta_{\bf k}
    +\tilde{P}_{-\bf k}\cos\theta_{\bf k}\right),
\end{eqnarray}
from the relation $(\ref{BnBt})$. We see this involves two steps: 1)
$(\tilde{Q}_{\pm\bf k}, \tilde{P}_{\pm\bf k})$ are first rotated locally in 
angles $\sigma_{\bf k}$ and $-\theta_{\bf k}$ on the ${+\bf k}$ and ${-\bf k}$ 
modes, respectively, and then 2) squeezed globally in squeeze parameter
$r_{\bf k}$. Local operations do not affect the entanglement measure:
The phases $\sigma_{\bf k}$ and $\theta_{\bf k}$ are not present in the 
reduced density matrix (RDM) transformed from $(\ref{RedWignerB})$, 
neither is any function of the RDM such as the entanglement entropy 
$(\ref{vNEn})$. Here only squeezing which is a global operation is relevant 
to quantum entanglement.

We now turn to the question of how to obtain the phase information and
whether/how it enter into entanglement dynamics considerations.

\subsection{Quantum phase}
\label{QPhase}

Mathematically the phase $\theta_{\bf k}$ of $\beta_{\bf k}$ in the
parametrization $(\ref{squeezpara})$ can be obtained by evaluating
\begin{equation}
  \theta_{\bf k} = -{i\over 2}
      \ln {\dot{\chi}_{\bf k}(\eta_0 )\chi^{}_{\bf k}(\eta )-
     \chi^{}_{\bf k}(\eta_0 )\chi'_{\bf k}(\eta ) \over
     \chi^*_{\bf k}(\eta_0 )\chi'^*_{\bf k}(\eta )-
     \dot{\chi}^*_{\bf k}(\eta_0 )\chi^*_{\bf k}(\eta)},
  \label{phaseB}
\end{equation}
and $\sigma_{\bf k}$ in $\alpha_{\bf k}$ can be obtained in a similar way.
But physically only the phase sum $\sigma_{\bf k} + \theta_{\bf k}$
could be measured. %The reason is the following.
%One hint of this is that from $(\ref{wavefnk})$ the probability density         %%%2/27
%$\Psi_0^* \Psi_0$ depends only on $| \chi_{\bf k}(\eta) |^2$, which is a          %
%function of $e^{\pm i(\sigma_{\bf k}+\theta_{\bf k})}$ only, while
%$\sigma_{\bf k} - \theta_{\bf k}$ was canceled since it is a global phase 
%in $\chi_{\bf k}$ from $(\ref{chiAsB})$ and $(\ref{squeezpara})$. 
%Below one will further see that the phase sum $\sigma_{\bf k}+\theta_{\bf k}$ 
%is actually the conjugate variable complementary to the particle number in        %
%our squeezed states.                                                            %%%

Transforming $(\ref{WignerB})$ to the density matrix in quadrature
amplitudes $Q^{}_{\pm\bf k}$ representation, one obtains 
\begin{equation}
  \rho[\cdots,Q^{}_{\bf k},Q^{}_{-\bf k},\cdots;\cdots, Q'_{\bf k},
  Q'_{-\bf k},\cdots] = \Psi[\cdots,Q^{}_{\bf k},Q^{}_{-\bf k},\cdots]
  \Psi^*[\cdots,Q'_{\bf k},Q'_{-\bf k},\cdots], 
  \label{DMQpQm}
\end{equation}
where $\Psi = {\prod_{\bf k}}^+ \Psi_{\bf k}$ and
\begin{eqnarray}
  \Psi_{\bf k} &=& {e^{-i\int^\eta d\bar{\eta}{\cal E}_0^{\bf k}(\bar{\eta})/\hbar}
    \over\sqrt{\pi}{\cal G}_{\bf k}^{1/4}}\exp{-1\over 2 {\cal G}_{\bf k}}
    \left\{ \left( 1+2n_{\bf k}+c^{}_{\bf k}{}^2-c^*_{\bf k}{}^2\right)\left(
    Q_{\bf k}^2 + Q_{-\bf k}^2\right) + 4\left[ c^{}_{\bf k}(n_{\bf k}+1)
    -c^*_{\bf k} n_{\bf k}\right]Q^{}_{\bf k}Q^{}_{-\bf k}\right\}
  \nonumber\\ &=& {e^{-i\int^\eta d\bar{\eta}{\cal E}_0^{\bf k}(\bar{\eta})/\hbar}
      \over \cosh r_{\bf k}}\left[
    {1+\sinh^2 r_{\bf k}\left(1-e^{2i(\sigma_{\bf k}+\theta_{\bf k})}\right)
      \over 1+\sinh^2 r_{\bf k}\left(1-e^{-2i(\sigma_{\bf k}+\theta_{\bf k})}
      \right)}\right]^{1/4}\sum_{n=0}^\infty \left( e^{i(\sigma_{\bf k}+
    \theta_{\bf k})} \tanh r_{\bf k}\right)^n
    \Phi_n\left(Q^{}_{\bf k}\right)\Phi_n\left(Q^{}_{-\bf k}\right),
\label{WavefnQ}
\end{eqnarray}
with ${\cal G}\equiv (1+2n_{\bf k})^2 -(c^{}_{\bf k} + c^*_{\bf k})^2$
and the number eigenstates in $Q$ representation,
\begin{equation}
   \Phi_n(Q) \equiv \sqrt{1\over 2^n n!\sqrt{\pi}}H_n(Q)e^{-Q^2/2}.
\end{equation}
So the density matrix $(\ref{DMQpQm})$ can be expressed as 
\begin{equation}
  \rho = {\prod_{\bf k}}^+ \sum_{n,m}\rho^{\bf k}_{nn,mm} 
  \Phi^{}_n\left(Q^{}_{\bf k}\right)\Phi^{}_n\left(Q^{}_{-\bf k}\right)
  \Phi^*_m\left(Q'_{\bf k}\right)\Phi^*_m\left(Q'_{-\bf k}\right),
  \label{DMQpQm1}
\end{equation}
with
\begin{equation}
  \rho^{\bf k}_{nn,mm} = {\tanh^{n+m} r^{}_{\bf k}\over 
    \cosh^2 r^{}_{\bf k}} e^{i(n-m)(\sigma^{}_{\bf k}+\theta^{}_{\bf k})}.
  \label{DMnm}
\end{equation}
Equation $(\ref{WavefnQ})$ shows that ${\bf k}$ and $-{\bf k}$ particles 
associated with $Q^{}_{\bf k}$ are always created in pairs, because the 
outcome of the measurement on numbers of ${\bf k}$ and $-{\bf k}$ particles 
separately will always be the same. From $(\ref{DMQpQm1})$ and according to 
\cite{PB97}, one can write down the probability distribution of the quantum 
phase-sum $\theta_+$,
\begin{equation}
  P(\theta_+) =\left\{2\pi\left[\cosh^2 r{}_{\bf k} + \sinh^2 r{}_{\bf k}
  - 2\cosh r{}_{\bf k}\sinh r{}_{\bf k}\cos \left(\theta_+ -\sigma^{}_{\bf k}
  -\theta^{}_{\bf k}\right)\right]\right\}^{-1},
\end{equation}
which peaks at $\theta_+ -( \sigma^{}_{\bf k} + \theta^{}_{\bf k}) = 2n\pi$,
$n\in{\bf Z}$, while the probability distribution of the quantum
phase-difference $\theta_-$, $P(\theta_-)= 1/2\pi$, is constant through
$0\le\theta_- < 2\pi$ so $\theta_-$ of the quantum state $(\ref{WavefnQ})$ is 
totally uncertain. This shows that $[(\sigma^{}_{\bf k}+\theta^{}_{\bf k})$ 
mod $2\pi]$ is the quantum phase complementary to the particle number,
while $\sigma_{\bf k}$ and $\theta_{\bf k}$ cannot be observed separately.

The matrix elements $(\ref{DMnm})$ are proportional to $c^{n-m}_{\bf k}$
\footnote{From here one can trace out $Q^{}_{-\bf k}$ and $Q'_{-\bf k}$ to get 
the RDM $\rho^{+\bf k}_{n,m} = \delta_{nm}\tanh^{2n} r_{\bf k}/\cosh^2 r_{\bf k}
\equiv \delta_{mn} \rho^{+\bf k}_{0,0} e^{-n\hbar\Omega/k_B T_{\rm eff}}$.
One could presumably identify an ``effective temperature" $T_{\rm eff} = 
\hbar\Omega/[k_B \ln (1+ n_{\bf k}^{-1})]$ depending only on $n_{\bf k}$ 
and valid for all values of $n_{\bf k}$, just like the vN entropy 
$(\ref{vNEn})$. Since such an effective temperature can vary quickly in 
time, it is very remote from the usual concept of temperature defined in 
statistical mechanics (see the discussion in Section \ref{EEnESM}).}.
If the phase changes so fast that $c^{}_{\bf k}$ could not be measured
precisely by any apparatus, then the information for quantum state
tomography will never be complete. Most likely in this case the off-diagonal
elements with $m\not= n$ would be averaged out, then the quantum state of
the field may appear like a classical state. This ``fake
decoherence" due to the technical limitation of measurement is different from
environment-induced decoherence such as from intermode couplings.

If the resolution of the apparatus gets higher, more phase information could 
then be observed. However, it is easy to verify that once $\rho^{\bf k}_{nn,mm} 
\propto c_{\bf k}^{n-m}$ for all nonzero $c^{}_{\bf k}\in{\bf C}$, the vN 
entropy of $\rho$ will be zero. So vN entropy cannot be generated by just 
replacing all the original $c^{}_{\bf k}$ by some outcomes of measurement 
with smaller absolute values. One simple way to produce vN 
entropy is to perform a truncation such as $\rho^{\bf k}_{nn,mm}\equiv 0$ for 
all $|n-m|> N$ with some positive integer $N$, meaning that the off-diagonal
elements oscillating quicker than $\exp\pm i N (\sigma^{}_{\bf k}+\theta^{}_{
\bf k})$ are not resolvable and being averaged out. For this truncated 
effective density matrix, the purity will be
\begin{equation}
  {\cal P}_{\rm eff} = {\prod_{\bf k}}^+ {1\over 1+ 2n^{}_{\bf k}}\left[ 1+ 
  2\sum_{m=1}^N\left( n^{}_{\bf k}\over 1+n^{}_{\bf k}\right)^m\right].
\label{purityN}
\end{equation}
For all $n_{\bf k}\ge 0$, the larger $N$ is, the closer ${\cal P}_{\rm eff}$ is 
to unity, so the effective density matrix is purer, and the vN entropy of it is 
closer to zero.

The RDM obtained by tracing out the $Q^{}_{-\bf k}$ and $Q'_{-\bf k}$ components 
in $(\ref{DMQpQm1})$ reads 
\begin{equation}
  \rho^R = {\prod_{\bf k}}^+ \sum_{n}\rho^{\bf k}_{nn,nn} 
  \Phi^{}_n\left(Q^{}_{\bf k}\right)\Phi^*_n\left(Q'_{\bf k}\right).
  \label{RDMk}
\end{equation}
One can immediately see that the vN entropy of $(\ref{RDMk})$, which is the 
entanglement entropy between the particles with $\bf k$ and $-\bf k$, has exactly 
the same value as the vN entropy of the effective density matrix of the vacuum 
with all off-diagonal elements averaged out, namely, 
\begin{equation}
  \rho^{}_{\rm eff} = {\prod_{\bf k}}^+ \sum_{n}\rho^{\bf k}_{nn,nn} 
  \Phi^{}_n\left(Q^{}_{\bf k}\right)\Phi^{}_n\left(Q^{}_{-\bf k}\right)
  \Phi^*_n\left(Q'_{\bf k}\right)\Phi^*_n\left(Q'_{-\bf k}\right).
  \label{DMeff}
\end{equation}
Thus one could say that the tracing-out process in obtaining the RDM 
$(\ref{RDMk})$ represents the ``strongest" coarse graining, though the 
coincidence of the entropy values here does not occur for general quantum 
states.

One should be careful that simply ignoring fast-oscillating
elements could create one more problem if our quantum state tomography is 
designed to reconstruct the Wigner function. The amplitude of $c_{\bf k}$ 
plays an important role in obtaining the correct entanglement entropy from 
the Wigner function. If one finds that $c_{\bf k}$ in $(\ref{WignerB})$ 
appears to be zero, then the factor in the exponent of the reduced Wigner 
function $(\ref{RedWignerB})$ will be $2(2n_{\bf k}+1)$ rather than $2/(
2n_{\bf k}+1)$, so that the effective reduced Wigner function 
%obtained from $(\ref{WignerB})$ with $c_{\bf k}$ terms dropped
cannot be transformed back to the correct effective RDM. This effective 
reduced Wigner function yields an incorrect vN entropy or purity for the 
${\bf k}$ particles, though here the vN entropy is no longer a 
well-defined entanglement entropy since the density matrix of the 
$({\bf k},-{\bf k})$ mode pairs (not the RDM of the ${\bf k}$ particles)
effectively constitutes a mixed state.
 
Since the quantum phase is conjugate to the particle number of a squeezed
state, one may expect that 
one could obtain the phase information from time derivatives of
$n_{\bf k}$ or equivalently, from evolution of the entanglement
entropy in time. Indeed, by noting that $\partial_{\eta_0}\chi^{}_{\bf k}
(\eta) = 0$ and from $(\ref{chiAsB})$, one has $\chi'_{\bf k}
(\eta) = \alpha^*_{\bf k}(\eta_0,\eta)\dot{\chi^{}_{\bf k}}(\eta_0) -
\beta^{}_{\bf k}(\eta_0,\eta)\dot{\chi^*_{\bf k}}(\eta_0)$.
This implies that
\begin{equation}
  n'_{\bf k} = -2\, {\rm Im}\left\{ c^{}_{\bf k}\left[
  \dot{\chi}^*_{\bf k}{}^2(\eta_0) + \Omega^2(\eta)
  \chi^*_{\bf k}{}^2(\eta_0)\right]\right\},
\label{Nvary}
\end{equation}
where $c^{}_{\bf k} \equiv \alpha^{}_{\bf k}\beta^{}_{\bf k}= -
e^{i(\sigma_{\bf k}+\theta_{\bf k})}\cosh r_{\bf k} \sinh r_{\bf k}$
provides information of the phase $\sigma_{\bf k} +\theta_{\bf k}$.
Unfortunately, in the right hand side of $(\ref{Nvary})$, $c^{}_{\bf k}$
is always multiplied by a term in the square bracket, which often cancels
the oscillation of $c^{}_{\bf k}$ so that one cannot read off the quantum
phase from the behavior of $n'_{\bf k}$.

Even $c^{}_{\bf k}$ {\it per se} are not always fast oscillating, though.
Below we will give an example when the off-diagonal elements of $\rho^{\bf
k}_{nn,mm}$ associated with the particle-number operators defined at the
initial moment do not oscillate, namely, when the Universe undergoes
inflationary expansion. But before we proceed, we want to make one more
remark on an alternative oscillating ``phase."

\subsection{Quantum Vlasov equation}

The $\eta$ time derivative of $c^{}_{\bf k}$ reads
\begin{equation}
  c'_{\bf k} = 2ic^{}_{\bf k}\left[\left|\dot{\chi^{}_{\bf k}}
    (\eta_0)\right|^2 + \Omega^2(\eta)\left|\chi^{}_{\bf k}(\eta_0)
    \right|^2\right]- i(2 n^{}_{\bf k}+1)\left[\left( \dot{\chi^{}_{\bf k}}
    (\eta_0)\right)^2 + \Omega^2(\eta)\left(\chi^{}_{\bf k}(\eta_0)
    \right)^2\right] \label{Cvary}
\end{equation}
Similar to \cite{KME98}, one can express $c_{\bf k}$ in terms of
$n_{\bf k}$ by solving ($\ref{Cvary}$) then insert it into
($\ref{Nvary}$) to get
\begin{eqnarray}
  & &n'_{\bf k}(\eta)= 2{\rm Re}\left\{\left[\dot{\chi}^{}_{\bf k}{}^2
    (\eta_0)+\Omega_{\bf k}^2(\eta)\chi^{}_{\bf k}{}^2(\eta_0)\right]
        \right.\times\nonumber\\& &\,\,\left.
    \int^\eta_{\eta_0} d\bar{\eta}\left( 2n_{\bf k}(\bar{\eta})+1\right)
    \left[ \dot{\chi}^*_{\bf k}{}^2(\eta_0)
      +\Omega_{\bf k}^2(\bar{\eta})\chi^*_{\bf k}{}^2(\eta_0)\right]
    e^{-2i[\Theta^{}_{\bf k}(\eta)-\Theta^{}_{\bf k}(\bar{\eta})]}
    \right\}, \label{Nfinal}
\end{eqnarray}
where
\begin{equation}
  \Theta^{}_{\bf k}(\eta)\equiv \int^\eta d\tau \left[
    \left| \dot{\chi}^{}_{\bf k}(\eta_0)\right|^2 + \Omega_{\bf k}^2
    (\tau)\left| \chi^{}_{\bf k}(\eta_0)\right|^2\right].
    \label{Thetadef}
\end{equation}
which demonstratively illustrates that the evolution of $n_{\bf k}$ is
in general non-Markovian. Only in the case that the phase $|\Theta^{}_{\bf 
k} (\eta)-\Theta^{}_{\bf k}(\bar{\eta})|$ grows rapidly in $\bar{\eta}-\eta$,
would the $\bar{\eta}$ integration be effective only around $\bar{\eta}
\approx \eta$, and the right-hand side of $(\ref{Nfinal})$ becomes local 
in time.

Anyway, solving the quantum Vlasov equation $(\ref{Nfinal})$,
or $(\ref{Nvary})$ and $(\ref{Cvary})$, is equivalent to solving
$\chi^{}_{\bf k}$ from $(\ref{eomchi})$ and then calculate $n^{}_{\bf k}$
and $c^{}_{\bf k}$, which is much simpler. We put $(\ref{Nvary})$ and
$(\ref{Cvary})$ here simply to show the relation between the particle
number and the phases. Often it is not economic to solve them directly.

We should emphasize that $\Theta^{}_{\bf k}$ in $(\ref{Thetadef})$ and the
counterpart in \cite{KME98}, which is the phase of the adiabatic mode
function, %rather than the phase of Bogoliubov coefficients, 
are different from the quantum phase $\sigma_{\bf k} +\theta_{\bf k}$ in 
general.

%%%%%%%%%%%%%%%%%%%%%%%%%%%%%%%%%%%%%%%%%%%%%%%%%%%%%%%%%%%%%%%%%%%%%%%%%%%%%
\section{Entanglement in cosmological particle creation}

\subsection{A real scalar field in the FRW universe}
\label{rsfFRW}

A real scalar field $\Phi$ with mass $m$ minimally coupled to a curved
spacetime with metric $g_{\mu\nu}$ is described by the action,
\begin{equation}
 S = \int d^4 x \sqrt{-g} \left[ -{1\over 2}\partial_\mu\Phi
   \partial^\mu\Phi - {m^2 \over 2}\Phi^2 \right].
   \label{Svarphi}
\end{equation}
We are working with a test-field condition where the gravitational field 
$g_{\mu\nu}$ is a given background, in this case, the FRW spacetime, 
with line element
\begin{eqnarray}
  ds^2 &=&  a(\eta)^2\left[ -d\eta^2 + {dr^2\over 1-\kappa r^2} +
          r^2 d\theta^2 + r^2\sin^2\theta d\phi^2\right],\nonumber\\
  &\equiv& a(\eta)^2\left[ -d\eta^2 + h_{ij}dx^idx^j\right],
\end{eqnarray}
where $\kappa =1,0,-1$ corresponds to closed, flat and open universes
respectively. In terms of the conformal time $\eta$ and the conformal
scalar field defined as
\begin{equation}
  X (x) \equiv a(\eta)\Phi (x) ,
\end{equation}
one can rewrite the field action $S$ as
\begin{equation}
  S = \int d\eta d^3 x \sqrt{h}\left[ {1\over 2}X'^2 -
    {1\over 2}\partial_i X \partial^i X +
    {1\over 2}\left( {a''\over a} -m^2 a^2\right)X^2\right]
  \label{act0}
\end{equation}
plus a surface term $-\int d^3 x X^2 a'/2a$. The field equation then reads
\begin{equation}
   X'' +\left(m^2a^2-{a''\over a}-\nabla^2\right)X =0.
\end{equation}

To better handle the spatial derivatives in action $S$, we
perform a transformation
\begin{equation}
  X (x) = {\sum_{\bf k}}' {\cal Y}_{\bf k}({\bf x})\phi^{}_{\bf k}(\eta),
\label{sumkold}
\end{equation}
where ${\sum_{\bf k}}'$ and ${\cal Y}_{\bf k}({\bf x})$ for $\kappa=-1$,
$0$, and $1$ can be found in \cite{BirDav}. 
For $\kappa=0$ (spatially flat), ${\sum_{\bf k}}'\equiv \int d^3 k/
(2\pi )^3$ and ${\cal Y}_{\bf k}$ is simply $e^{i{\bf k}\cdot{\bf x}}$,
and the action written in the Fourier $k$ space becomes $(\ref{SphiRgen})$
with time-varying squared frequencies
\begin{equation}
  \Omega_{\bf k}^2(\eta) \equiv k^2 + m^2 a^2
  -{a''\over a}. \label{wketa}
\end{equation}
So the formulation in Sec. \ref{PhiR} can be directly applied.

%%%%%%%%%%%%%%%%%%%%%%%%%%%%%%%%%%%%%%%%%%%%%%%%%%%%%%%%%%%%%%%%%%%%%%%
\subsection{Entanglement of particle creation in a de Sitter spacetime}

In the spatially flat FRW coordinatization of the de Sitter space, the scale
factor is $a = -(H\eta)^{-1} = e^{H t}$ with Hubble constant $H$ and
cosmic time $t$, such that $\eta$ runs from $-\infty$ to $0$ as $t$ goes
from $-\infty$ to $\infty$ (see, for example, Sec. 5.4 in Ref.\cite{BirDav}).
The squared time-varying natural frequency in $(\ref{wketa})$ now reads
\begin{equation}
  \Omega^2_{\bf k} = k^2+\left({m^2\over H^2} -2\right){1\over \eta^2}.
\end{equation}
If $m^2 < 2H^2$, $\Omega^2_{\bf k}$ will become negative at late times
($\eta\to 0$) for all finite $k$. The Bunch-Davies vacuum corresponds to 
taking the value \cite{BunDav78}
\begin{equation}
  \chi^{}_{\bf k} = \sqrt{\pi \eta \over 4}H_\nu^{(2)}(k\eta),
  \label{dSmode}
\end{equation}
where $H_\nu^{(2)}$ is the Hankel function and $\nu \equiv [(9/4)-
(m^2/H^2)]^{1/2}$. 

The phase in the quantum Vlasov equation $(\ref{Nfinal})$ of such a scalar
field in this de Sitter spacetime
\begin{equation}
  \Theta^{}_{\bf k}(\eta)-\Theta^{}_{\bf k}(\bar{\eta}) = (\eta-\bar{\eta})
    \left\{ \left|\dot{\chi}^{}_{\bf k}(\eta_0)\right|^2 + \left[ k^2+\left(
    {m^2\over H^2}-2\right){1\over\eta\bar{\eta}}\right]\left|\chi^{}_{\bf k}
    (\eta_0)\right|^2\right\}_{\eta_0\to -\infty},
\end{equation}
varies quite rapidly in $\eta-\bar{\eta}$ time, so the integration in
$(\ref{Nfinal})$ is more pronounced around $\bar{\eta}\approx \eta$ and
the behavior of $n_{\bf k}$ in an inflationary universe can be treated
in a Markovian approximation. However, this does not imply that the
quantumness of the field is lost in this epoch, as some authors working
on the decoherence of quantum fields in inflationary cosmology are drawn
to making such a claim.

Substituting $(\ref{dSmode})$ into $(\ref{phaseB})$ and the counterpart for
$\sigma_{\bf k}$, we find that $\sigma_{\bf k}+\theta_{\bf k}$ varies quite
slowly except when $\eta^2 \approx \left|[(m^2/H^2)-2]/k^2\right|$, which is 
around the moment that $\Omega_{\bf k}^2$ is crossing zero in the cases 
with $m^2 < 2H^2$. Far from this moment in cosmic time $t$, $c_{\bf k}^{n-m}$ 
almost does not oscillate and could be identified clearly for very large 
$|n-m|$. Therefore perhaps contrary to common belief, in almost the whole 
inflation epoch, the off-diagonal elements of the density matrix corresponding 
to the quantum interference between the ${\bf k}$ and $-{\bf k}$ particles 
associated with $B^{}_{\bf k}$ corresponding to the in-vacuum manifest and
the coarse-grained effective density matrix is extremely pure \footnote{
Campo and Parentani have come to a similar conclusion in \cite{CP08b} for a 
single interacting field, rather than the free field in this paper, in 
inflationary universe.}.

Alternatively, in the adiabatic number basis, the two concepts of phase in 
Sec. $\ref{phaseinf}$ can coincide. For example, the (first-order) 
adiabatic mode function used by KME in \cite{KME98} reads
\begin{equation}
  \tilde{\chi}_{\bf k}(\eta) = \sqrt{1\over 2\Omega_{\bf k}(\eta)}
    \exp -i\int^\eta_{\eta_0} d\tilde{\eta} \Omega_{\bf k}(\tilde{\eta}).
  \label{WKBmode}
\end{equation}
Numerically we find that at early times when $\Omega_{\bf k}$ is real 
and not very small, the adiabatic particle number \cite{KME98}
\begin{equation}
  {\cal N}_{\bf k} = \left|\tilde{\chi}_{\bf k} \left( \chi'_{\bf k}
    + i\Omega_{\bf k}\chi^{}_{\bf k}\right)\right|^2
\end{equation}
is indeed much less than $n_{\bf k}$, and the counterpart of
$c^{}_{\bf k}$ for the adiabatic mode function does oscillate,
in exactly the same way as the oscillation in the phase $\Theta$ in their 
quantum Vlasov equation (which implies that it is still impossible
to determine the quantum phase by observing the evolution of their
${\cal N}_{\bf k}$). This justifies the argument in \cite{KME98}: The 
adiabatic particles look more classical since the off-diagonal elements of 
the density matrix of the vacuum (the counterpart of $(\ref{DMnm})$ 
with $r^{}_{\bf k}$ and $\sigma^{}_{\bf k}+\theta^{}_{\bf k}$ obtained in 
the adiabatic number basis) oscillate too fast to be resolved. 
Then the vN entropy ${\cal S}_{\rm eff}$ of the effective density matrix 
$\rho^{}_{\rm eff}$ of the vacuum with all off-diagonal elements in the 
adiabatic number representation vanishing is $(\ref{vNEn})$ with 
$n^{}_{\bf k}$ replaced by ${\cal N}_{\bf k}$ [cf. Eq.(3.23) in 
\cite{KME98} for complex scalar fields],
\begin{equation}
  {\cal S}_{\rm eff}= -{\rm Tr}\rho^{}_{\rm eff} \ln\rho^{}_{\rm eff}=
  {1\over 2}{\sum_{\bf k}}\left[({\cal N}_{\bf k}+1)\ln ({\cal N}_{\bf k}+1) 
    -{\cal N}_{\bf k}\ln {\cal N}_{\bf k}\right], \label{Seffadia}
\end{equation}
which is also valid for all values of ${\cal N}_{\bf k}$ and is much less 
than the vN entropy in $n_{\bf k}$. Again, as discussed in Sec.
\ref{QPhase}, the value of the above $S_{\rm eff}$ is the same as the value 
of the entanglement entropy between the adiabatic particles with ${\bf k}$ and 
$-{\bf k}$, while $S_{\rm eff}$ will decrease as more and more phase 
information is resolved and the off-diagonal elements manifest.

Unfortunately $(\ref{WKBmode})$ is not well defined if $\Omega_{\bf k}
=0$, which occurs in the case $m^2/H^2 <2$ when the physical
wavelength of the mode crosses the horizon. At that very moment,
${\cal N}_{\bf k}$ diverges and the WKB approximation $(\ref{WKBmode})$
fails. In this case, after $|\eta| = (2-(m^2/H^2))/k^2$, when the
wavelength of the field mode is longer than the size of the horizon
(superhorizon), the notion of adiabatic particle is no longer viable until 
the inflation era ends and the universe becomes radiation dominated.

%%%%%%%%%%%%%%%%%%%%%%%%%%%%%%%%%%%%%%%%%%%%%%%%%%%%%%%%%%%%%%%%%%%%%%%
\section{Summary Remarks}

We conclude with two remarks pointing to the main themes stated in the
Introduction, namely, quantum entanglement depends on partition as well
as the choice of physical variables or measurables. They pertain to 1)
the conditions of separability of quantum states in particle creation
processes and 2) the relation of entanglement dynamics and entropy
generation both measured by the von Neumann entropy.

\subsection{Conditions of separability of quantum states}

For the model of a free scalar field theory in a dynamical background
field or spacetime we see clearly that different quantities imply
different separabilities of quantum states in different partition and
different measurables. Using the vN entropy as a common
currency for comparisons, we see that

1. The vN entropy of $\rho ={\prod_{\bf k}}^+ \rho^{\bf k}= {\prod_{\bf k}}^+ 
\Psi^{}_{\bf k} \Psi^*_{\bf k}$ with $(\ref{WavefnQ})$ for $({\bf k},
-{\bf k})$ mode pair, or of the Wigner functions $(\ref{WignerBt})$ or 
$(\ref{WignerB})$, vanishes. This means that $\rho$ is a pure state, and the 
quantum field here is a completely isolated system even in a classical 
dynamical background field or spacetime. The vanishing vN entropy of 
$(\ref{rhoAtBt})$ and $(\ref{rhoAB})$ is similar.

2. The vN entropy of ${\rm Tr}_{{\bf p} \not= \pm{\bf k}}\,\{\rho \}$
for some specific ${\bf k}$ with $\rho = {\prod_{\bf k}}^+ \Psi^{}_{\bf k} 
\Psi^*_{\bf k}$, $(\ref{WignerBt})$, or $(\ref{WignerB})$, is zero. 
This means that each $({\bf k}, -{\bf k})$ mode pair is separable from other 
mode pairs, and of course, each $\Psi_{\bf k}$ is pure.

3. The vN entropy of the reduced Wigner function obtained by tracing out, 
say, the $k^3<0$ components of $(\ref{WignerBt})$, is zero. This means that 
the quantum state of $\tilde{B}^{}_{\bf k}\tilde{B}^*_{\bf k}$ is separable 
from $\tilde{B}^{}_{-\bf k}\tilde{B}^*_{-\bf k}$ and all other field modes, 
while no particle with $\pm{\bf k}$ associated with the particle counter 
or the number operator defined by $b^{}_{\bf k}(\eta)$ and $b^\dagger_{\bf k}
(\eta)$ is created. The zero vN entropy of the reduced Wigner function
from $(\ref{rhoAtBt})$ is similar.                         

4. The vN entropy $(\ref{vNEn})$ of $(\ref{RedWignerB})$ is nonzero, 
meaning that the particles with ${\bf k}$ associated with the particle 
counter defined by $b^{}_{\bf k}(\eta_0)$ and $b^\dagger_{\bf k}(\eta_0)$
%, which are physical,
are entangled with their $-{\bf k}$ partners.
The nonzero vN entropy of $(\ref{redrhoC})$ has a similar meaning.

5. The vN entropy of the exact density matrix of the vacuum $\rho$ is zero, 
but the vN entropy of the effective density matrix in Fock representation 
with off-diagonal elements averaged out is not. The latter indicates that 
the phase information has been coarse grained.

6. The value of the vN entropy of the effective density matrix $(\ref{DMeff})$ 
with all off-diagonal elements averaged out [see statements above $(
\ref{purityN})$] is the same as the value of the vN entropy of the reduced 
density matrix $(\ref{RDMk})$ after tracing out the particles with $-\bf k$.
This suggests that the tracing-out process in obtaining the RDM represents
the strongest coarse graining. 

7. In the adiabatic number basis, the vN entropy $(\ref{Seffadia})$ describing 
the entanglement between the adiabatic particles with ${\bf k}$ and their 
$-{\bf k}$ partners has a different value from $(\ref{vNEn})$, due to a 
different choice of physical measurables.

\subsection{Entanglement entropy and entropy in statistical mechanics}
\label{EEnESM}

As we saw in the above the von Neumann entropy has also been used as
a measure of the entropy generation in particle creation processes.
In \cite{KME98} KME argued that the effective density matrix
$\rho^{}_{\rm eff}$ would appear as a mixed state when the off-diagonal
elements oscillate too rapidly to be resolved. We see close similarity
between the vN entropy of a bosonic field used in this context of
nonequilibrium mechanics and that measuring the quantum entanglement
between the $({\bf k}, -{\bf k})$ particles in a single mode pair.
Indeed, from our result $(\ref{vNEn})$ we see that the correlation
between each particle pair seems to be equal, so that the entanglement
entropy between $\bf k$ and $-{\bf k}$ particles seems to be counting
the number of the degrees of freedom that are integrated out. 
Although the entropy of the former has the same value as the latter
in the cases considered in this paper, the differences between these two 
forms of entropy are perhaps more revealing, especially when viewed from 
their respective theoretical structures.

Let us compare the difference between the vN entropy $S=-{\rm Tr}\rho
\log\rho$ \cite{LL80} of a closed quantum system with the Boltzmann's
entropy in a microcanonical ensemble $S =-k_B\log \Omega$ where $\Omega$ 
is the number of accessible states.  Note that both describes an isolated
quantum system. The enumeration of accessible states is independent of
the representation and can contain both entangled states and separable
states.  When one assumes that $\Omega= {\rm Tr} \rho$ is given by the
probability ${\rm Tr} \rho$ of finding an isolated system in a
particular quantum state, one has already ignored all physical
information contained in the off-diagonal components of the density
matrix such as quantum phase of quantum states. This is accomplished
under the random phase approximation whereby one can use the
concepts of probabilities exclusively to describe all statistical
mechanical properties of the system.  When a system can occupy all of
its accessible states with equal {\it a priori} probability, then the 
system is in equilibrium. These are, as we know, the two fundamental
postulates of equilibrium statistical mechanics, namely, (1) equal {\it a
priori} probability, and (2) random phase \cite{Huang}. (For a depiction
of how a quantum system in contact with a thermal bath turn from a
quantum fluctuation dominated phase to a thermal fluctuation dominated
phase at increasing temperature,  and under what conditions will these
two postulates be satisfied in an open quantum system, see
\cite{HuZhangUncer}.) Finally it is when the particle number $n_{\bf k}\gg 
1$ that one can begin to use thermodynamic arguments. 

Thus we see the three stages distinctly:  vN and Boltzmann entropy 
(in its original form) both describe fully isolated quantum systems. When 
one begins to use probability for the description of the system, quantum 
phase information is lost. When one imposes in addition the equal {\it a 
priori} probability assumption one reaches an equilibrium condition.
Thermodynamic description requires an additional assumption that
both the number of particles and the volume of the system approach
infinity while their ratio is kept a constant. As is known and shown
in some of our earlier work \cite{HuTimeasy} the thermodynamic entropy
is different from the (equilibrium) statistical mechanical entropy and
the quantum (nonequilibrium) vN entropy, in increasing order of specificity.

In contrast, 
it is clear that the von Neumann entropy of one of the two parties of an 
isolated, bipartite system is a well-defined measure of entanglement 
\cite{BZ06} and $(\ref{vNEn})$ is a good entanglement entropy for all 
$n_{\bf k}$ for all time. \\

%In this paper we have investigated some basic theoretical issues for
%the analysis of entanglement dynamics in particle creation processes.
%In our follow-up work we will apply these results to the study of
%quantum entanglement in quantum processes involving parametric
%amplification (or squeezing) which are quite commonly encountered in
%quantum optics, and to the study of quantum entanglement and decoherence
%issues in cosmological particle creation with an eye toward observational
%consequences. \\

\noindent {\bf Acknowledgments}
SYL wishes to thank Kin-Wang Ng for illuminating discussions.
He also thanks the Institute of Physics, Academia Sinica for hospitality
during the development of this work.  BLH wishes to thank the hospitality
of NCTS and the QIS group at National Cheng Kung University of Taiwan.
This work is supported in part by  NSF Grants No. PHY-0426696 and No. 
PHY-0801368.

\end{document}